\documentclass[10pt,aps,prx,amsfonts,amsmath,amssymb,raggedbottom,longbibliography,reprint,superscriptaddress,citeautoscript]{revtex4-2}
\usepackage[usenames,dvipsnames]{color}
\usepackage{graphicx,microtype}
\usepackage{sidecap}
\usepackage[bookmarks=false,colorlinks]{hyperref}
\hypersetup{linkcolor=Plum,citecolor=Plum,filecolor=Plum,urlcolor=PineGreen}

\begin{document}

\title{Leggett Modes Accompanying Crystallographic Phase Transitions}

\author{Quintin N. Meier}
\affiliation{Université Grenoble Alpes, CEA, LITEN, 17 rue des Martyrs, 38054 Grenoble, France}

\author{Daniel Hickox-Young}
\affiliation{Department of Materials Science and Engineering, Northwestern University, Evanston, IL 60208, USA}

\author{Geneva Laurita}
\affiliation{Department of Chemistry and Biochemistry, Bates College, Lewiston, ME 04240, USA}

\author{Nicola A. Spaldin}
\affiliation{Materials Theory, ETH Zurich, Wolfgang-Pauli-Strasse 27, 8093 Z\"urich, Switzerland}

\author{James M. Rondinelli}
\affiliation{Department of Materials Science and Engineering, Northwestern University, Evanston, IL 60208, USA}
\affiliation{Northwestern-Argonne Institute of Science and Engineering (NAISE), Northwestern University, Evanston, IL 60208, USA}

\author{Michael R. Norman}
\email{norman@anl.gov}\affiliation{Materials Science Division, Argonne National Laboratory, Lemont, IL  60439, USA}

\date{\today}

\begin{abstract}
Higgs and Goldstone modes, well known in high energy physics, have been realized in
a number of condensed matter physics contexts, including superconductivity and magnetism.
The Goldstone-Higgs concept is also applicable to and gives rise to new insights on structural phase transitions.
Here, we show that the Leggett mode, a collective mode observed in multi-band superconductors, also has an analog
in crystallographic phase transitions.  Such structural Leggett modes can occur in the phase channel as in the original work of Leggett, \href{https://doi.org/10.1143/PTP.36.901}{Prog.\ Theor.\ Phys.\ \textbf{36}, 901 (1966)}.  That is, they are antiphase Goldstone modes (anti-phasons).
In addition, a new collective mode can also occur in
the amplitude channel, an out-of phase (antiphase) Higgs mode, that should be observable in multi-band superconductors as well.  We illustrate the existence and properties of these structural Leggett modes using the example of the pyrochlore relaxor 
ferroelectric, Cd$_2$Nb$_2$O$_7$.
\end{abstract}

\maketitle
\section{Introduction}
Spontaneous symmetry breaking is a ubiquitous phenomenon in physics, and is responsible for various collective
modes, most famously the well-known Goldstone and Higgs modes. The former refers to fluctuations of the phase of an order parameter,  the latter to fluctuations of its amplitude.  These modes play a fundamental role in gauge theories of particle physics, and are also important in condensed matter physics.
The Goldstone-Higgs
phenomenon in high energy physics is a relativistic generalization of the analogous behavior found in superconductors \cite{phil}. In superconductors, the Goldstone mode does not occur at zero energy as in a neutral superfluid, but it is pushed to the plasma frequency by coupling to the electromagnetic field \cite{phil2}.  The superconducting Higgs mode is non-trivial to observe since its energy tends to be located near the quasiparticle continuum \cite{schmid}.
Nevertheless, it has been observed in several superconductors  \cite{pekker}.

Superconductors can exhibit other collective modes.  In particular, a relative phase mode of the order parameters of the two bands of a two-band superconductor was proposed by Leggett \cite{leggett} and first realized in MgB$_2$ \cite{mgb2}. While the Goldstone mode is pushed to the plasma frequency by coupling to the electromagnetic field, the Leggett mode maintains charge neutrality and so is not affected by the field \cite{leggett}.  Other modes are known as well, for instance Carlson-Goldman modes in which the superconducting and normal condensates oscillate out of phase \cite{cg}. In addition, in superfluid $^3$He a variety of clapping, flapping, and Higgs modes occur because of the high degeneracy of its SO(3) $\times$ SO(3) $\times$ U(1) order parameter space \cite{wolfle,volovik}.

Structural phase transitions are typically associated with soft phonons \cite{scott}, and the concepts of Goldstone and Higgs modes have provided new insights into these transitions. A classic example is
the pyrochlore Cd$_2$Re$_2$O$_7$.  Its structural phase transition arises from an instability associated with a doubly degenerate $\Gamma_3^-$ phonon \cite{sergienko}. This defines a Mexican-hat free energy surface with the top of the hat representing the high temperature cubic phase, and the brim representing the lower symmetry distorted phase.  In the Landau free energy, anisotropy terms that warp the brim of the Mexican hat only arise at sixth order, suggesting the existence of a low energy Goldstone mode corresponding to oscillations along the brim. (Since they are not at zero energy because of the warping terms, these are sometimes referred to as pseudo-Goldstone modes).
Subsequent Raman scattering experiments \cite{kendziora} exhibited strong evidence for this mode, and its existence has been recently
confirmed by diffuse scattering studies \cite{venderley}.  The Higgs mode is also evident from the Raman data, though its
interpretation has been challenged by recent pump-probe measurements \cite{harter}.
After these pioneering studies of Cd$_2$Re$_2$O$_7$, Higgs and Goldstone modes have been proposed in a variety of perovskite and other complex oxides \cite{serge,singh,marthinsen,prosandeev,meier,zhang}.
Routes to detect them if they are optically silent  have also been identified \cite{juraschek}. 

Here, we investigate whether the Leggett mode can be realized in the structural context, where it could give new insights into the associated structural phase transitions.  We begin by developing a minimal Landau model that describes a structural phase transition that is accompanied by a new collective mode,  the antiphase Higgs mode (the amplitude analog of the Leggett mode), that should also be observable in multi-band superconductors.  We then consider a Landau treatment of the ferroelectric transition in the pyrochlore Cd$_2$Nb$_2$O$_7$, where we show that both the antiphase Higgs mode and the Leggett mode (the anti-phason) are possible.  Finally,
we discuss secondary modes that can drive these Leggett
modes by coupling to them in the context of pump-probe studies.
We conclude by discussing the relevance of our work to other materials, in particular those that involve modes at non-zero wavevectors.

\section{Minimal Landau Model}
A superconductor is described by an order parameter $\Delta e^{i\phi}$ with an amplitude, $\Delta$, and a phase, $\phi$.  In a two-band superconductor, one can identify a collective mode, called the Leggett mode, that corresponds to oscillations of the relative phase of the order parameters associated with the two bands, $\phi_1-\phi_2$ \cite{leggett}.  This mode is an eigenvector of a secular matrix whose eigenvalues are the collective mode energies \cite{sharapov}.  In the structural
context, the corresponding secular matrix is the force-constant matrix \cite{meier}, with elements $\varphi_{ij}=\frac{\partial^2F}{\partial u_i \partial u_j}$ where $u_i$ is the displacement
of the $i$-th ion from its high symmetry position and $F$ is the free energy. 
For a periodic system with $N$ atoms in the unit cell, this is a $3N \times 3N$ matrix.
For our further analysis, we will use a reduced form of the force-constant matrix, $\Phi_{ij}=\frac{\partial^2 F}{\partial q_i \partial q_j}$, where $q_i$ are symmetry-adapted distortion modes \cite{amp,iso}.  Each $q_i$ is a linear combination of atomic displacements that transform like a particular group representation of the high symmetry phase, and as such describe a collective motion involving multiple ions.
For structural phase transitions following  Landau theory \cite{cowley}, $F$
is formulated as a polynomial expansion in these $q_i$. The distortions from the high-symmetry phase are given, in the harmonic limit, by the eigenvectors of the force-constant matrix 
\footnote{%
We note that the force-constant matrix is related to, but not equal to, the dynamical matrix determining the phonons, whose elements instead are $\frac{\Phi_{ij}}{\sqrt{M_iM_j}}$
where $M_i$ is the mass of the $i$-th ion \cite{meier}. As such, a given Higgs or Goldstone mode, defined in terms of the eigenvectors of the force-constant matrix, can be an admixture of several phonons of the appropriate symmetry \cite{meier}.
}.
The advantage of Landau theory is the reduction of the large force-constant matrix to this smaller one involving only the $q_i$ relevant to the phase transition \cite{meier}.

To illustrate the structural Leggett mode, we present a simple example in which the phase transition involves only two modes, $\mathbf{q}$ and $\mathbf{r}$,
each from a different two-dimensional group representation.
For simplicity, we use a Cartesian basis, $\mathbf{q}~(q_1,q_2)$ and $\mathbf{r}~(r_1,r_2)$.  In a polar
basis, $q_i=Q_i\cos(\phi_i)$, etc.  In general, in the uncoupled case, each mode would have a different transition temperature, $T_\mathbf{q} \neq T_\mathbf{r}$. We describe each mode by a Mexican-hat potential, and include a biquadratic coupling between them \cite{holo}.
In this case, the free energy is given by
\begin{align}
    F &=\dfrac{a_\mathbf{q}}{2}(q_1^2+q_2^2)+\dfrac{b_\mathbf{q}}{4}(q_1^2+q_2^2)^2+\dfrac{a_\mathbf{r}}{2}(r_1^2+r_2^2)\nonumber\\
    &+\dfrac{b_\mathbf{r}}{4}(r_1^2+r_2^2)^2+\dfrac{c}{2}(q_1^2+q_2^2)(r_1^2+r_2^2)\,,
\label{eq:bq}
\end{align}
where the temperature dependence is typically only included in the quadratic terms $a_\mathbf{q}\equiv a_{\mathbf{q}0}(T-T_\mathbf{q})$ and $a_\mathbf{r}\equiv a_{\mathbf{r}0}(T-T_\mathbf{r})$.
The subspace of the force-constant matrix of interest is now given by
\begin{equation}
\Phi_{ij}(T)=\left.\dfrac{\partial^2 F}{\partial \phi_i\partial \phi_j}\right|_{\phi_i=\left<\phi_i\right>, \phi_j=\left<\phi_j\right>}
\end{equation}
where $\phi_i\in \{q_1,q_2,r_1,r_2\}$, and $\left<~\right>$ describes the thermodynamic average at a given temperature.

Choosing without loss of generality that $q_1=\left<q\right>$, $q_2=0$, $r_1=\left<r\right>$, $r_2=0$, the force-constant matrix then becomes
\begin{equation}
     \Phi=   \begin{bmatrix}
    H_\mathbf{q} & 2c\left<q\right>\left<r\right> & 0 & 0 \\
  2c\left<q\right>\left<r\right> & H_\mathbf{r} & 0 &  0 \\
   0 & 0 & G_\mathbf{q} & 0 \\
  0 & 0 & 0 & G_\mathbf{r}
\end{bmatrix} \quad ,
\end{equation}
with
\begin{align*}
    H_\mathbf{q}&=a_{\mathbf{q}0}(T-T_\mathbf{q})+3b_\mathbf{q}\left<q\right>^2+c\left< r\right>^2\\
    G_\mathbf{q}&=a_{\mathbf{q}0}(T-T_\mathbf{q})+b_\mathbf{q}\left<q\right>^2+c\left< r\right>^2\\
    H_\mathbf{r}&=a_{\mathbf{r}0}(T-T_\mathbf{r})+3b_\mathbf{r}\left<r\right>^2+c\left< q\right>^2\\
    G_\mathbf{r}&=a_{\mathbf{r}0}(T-T_\mathbf{r})+b_\mathbf{r}\left<r\right>^2+c\left< q\right>^2 \quad.
\end{align*}
Here, we use $H$ to indicate Higgs modes, that is amplitude modes with symmetry $A_1$, and $G$ to indicate Goldstone modes, that is phase modes whose symmetry depends on the underlying space groups involved.  

\begin{figure}
\centering
\includegraphics[width=0.95\columnwidth]{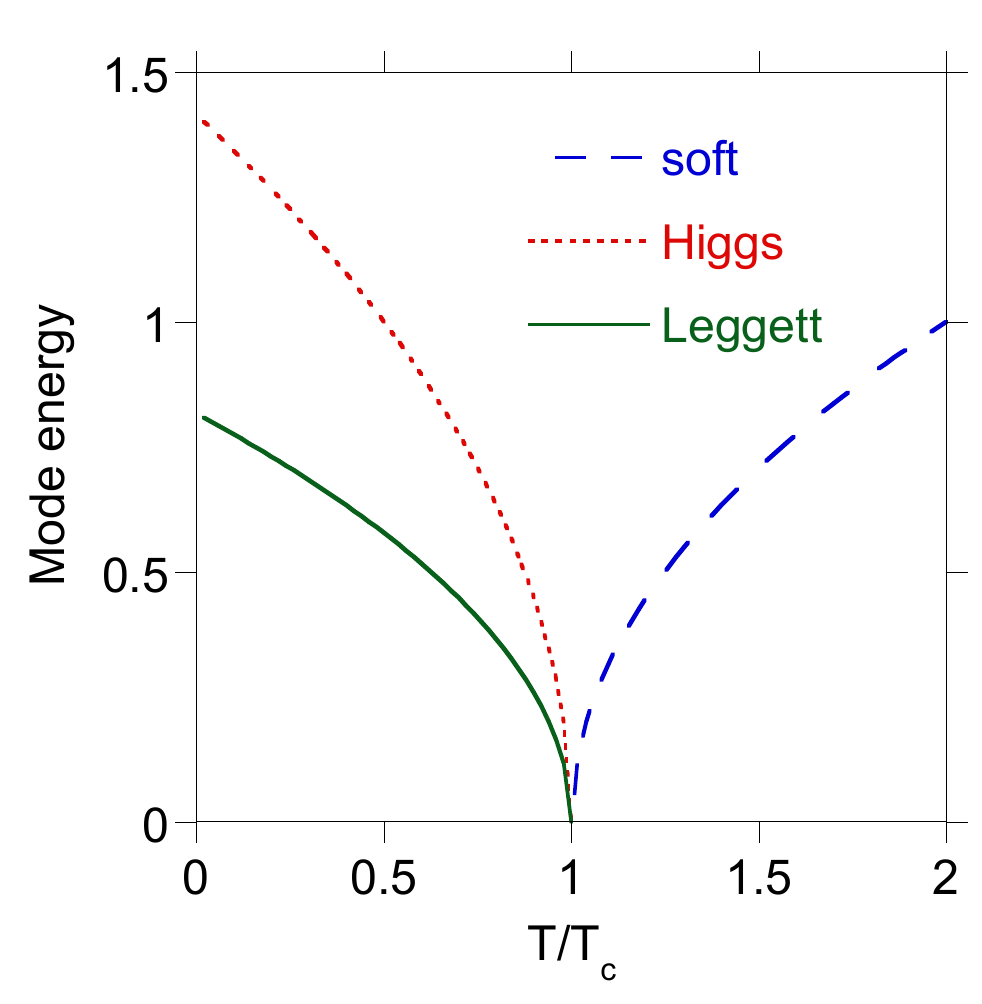}
\caption{Mode energies for the minimal Landau model with $a_\mathbf{q}=a_\mathbf{r}$, $b_\mathbf{q}=b_\mathbf{r}$, and $c=b_\mathbf{q}/2$, so that $T_\mathbf{q}=T_\mathbf{r}=T_\mathbf{c}$.  Units are such that $a_{\mathbf{q}0}/T_\mathbf{q}$ is set to unity.  The Goldstone modes (not shown) are at zero energy.}
\label{figy}
\end{figure}

\begin{SCfigure*}
\centering
\includegraphics[width=1.55\columnwidth]{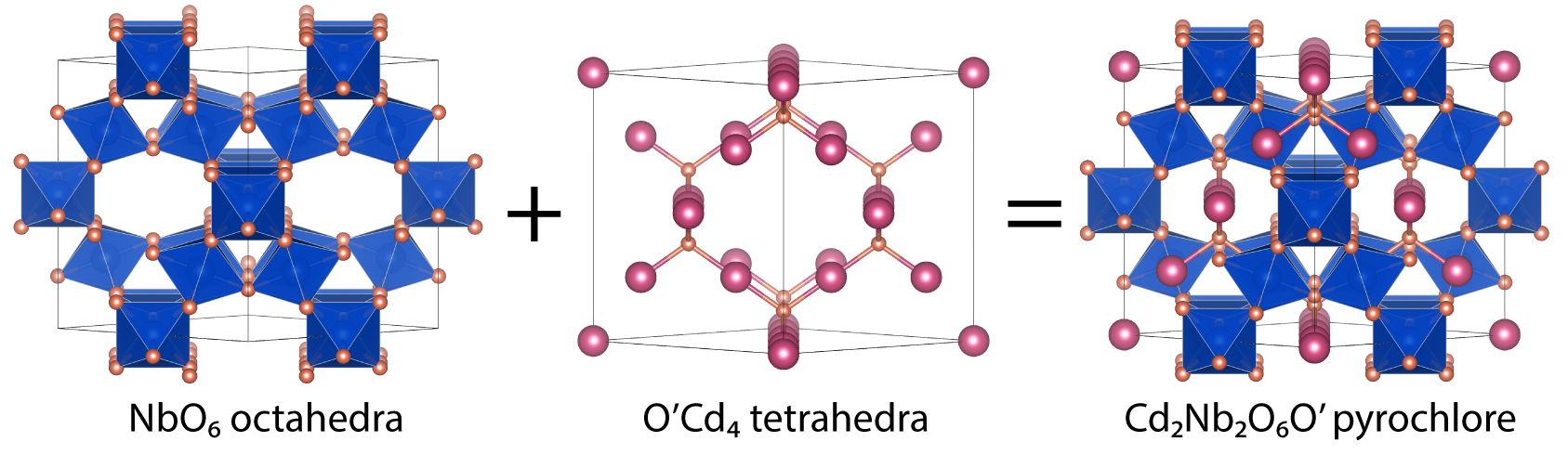}
\caption{Crystal structure of cubic  Cd$_2$Nb$_2$O$_7$ shown to 
illustrate its two interpenetrating sublattices of NbO$_6$ octahedra and O$^\prime$Cd$_4$ tetrahedra with Cd (magenta), Nb (blue), and O (orange) ions.

}
\label{figx}
\end{SCfigure*}

There are no
off-diagonal terms coupling the Higgs and Goldstone sectors, but in this simple example,
the biquadratic coupling term couples the two Higgs modes, $H_\mathbf{q}$ and $H_\mathbf{r}$.  Because of this coupling, we expect that a new collective mode,  the antiphase Higgs (the amplitude analog of the Leggett mode), can exist.
To see this, note that the eigenvalues of the force-constant
matrix are the square of the collective mode frequencies
\footnote{The $q_i$ are typically given in length units, meaning an effective mass is needed to convert the force-constant eigenvalues to frequency units.}.
That is, the eigenvalues of the upper $2\times2$ block of the force-constant matrix are
\begin{align}
    \omega^2_{\pm} =\dfrac{1}{2}\left(H_\mathbf{r}+H_\mathbf{q} \pm \sqrt{(H_\mathbf{r}-H_\mathbf{q})^2+16c^2\left<q\right>^2\left<r\right>^2}\right)\,.
\end{align}
 The eigenvectors are \begin{equation}
    (\mathbf{u},\mathbf{v})_{\pm}= \frac{1}{N_{\pm}}\left(1,\dfrac{\omega_{\pm}^2-H_\mathbf{q}}{2c\left<q\right>\left<r\right>}\right)\,,
    \label{eq:eigs}
\end{equation}
where $N_{\pm}$ is a normalization factor.
For small $c\langle q \rangle \langle r \rangle$ relative to $H_\mathbf{q}-H_\mathbf{r}$, we can think in terms of separate Higgs modes that are weakly coupled. In the limit of large coupling $c\langle q \rangle \langle r \rangle$, however, we obtain an in-phase Higgs mode and an out-of-phase Higgs mode.  The latter is the Higgs analog of the Leggett phase mode.  These are illustrated in \autoref{figy} for the  simple case where the $\mathbf{q}$ and $\mathbf{r}$ parameters are the same.  For this case, it is easy to show that the square of the Higgs mode energy is $-2a_\mathbf{q}$ and that of the Leggett mode is  $-2a_\mathbf{q}\frac{b_\mathbf{q}-c}{b_\mathbf{q}+c}$.

For its realization, one finds from the above equations that even one-dimensional group representations would produce this antiphase Higgs mode.  To determine the mode dispersions requires the addition of gradient terms
that typically lead to a quadratic dispersion about the ordering wavevector, unless the mode energy at the ordering vector is at zero energy, in which case the dispersion is linear in momentum instead.

\begin{figure*}
\centering
\includegraphics[width=1.3\columnwidth]{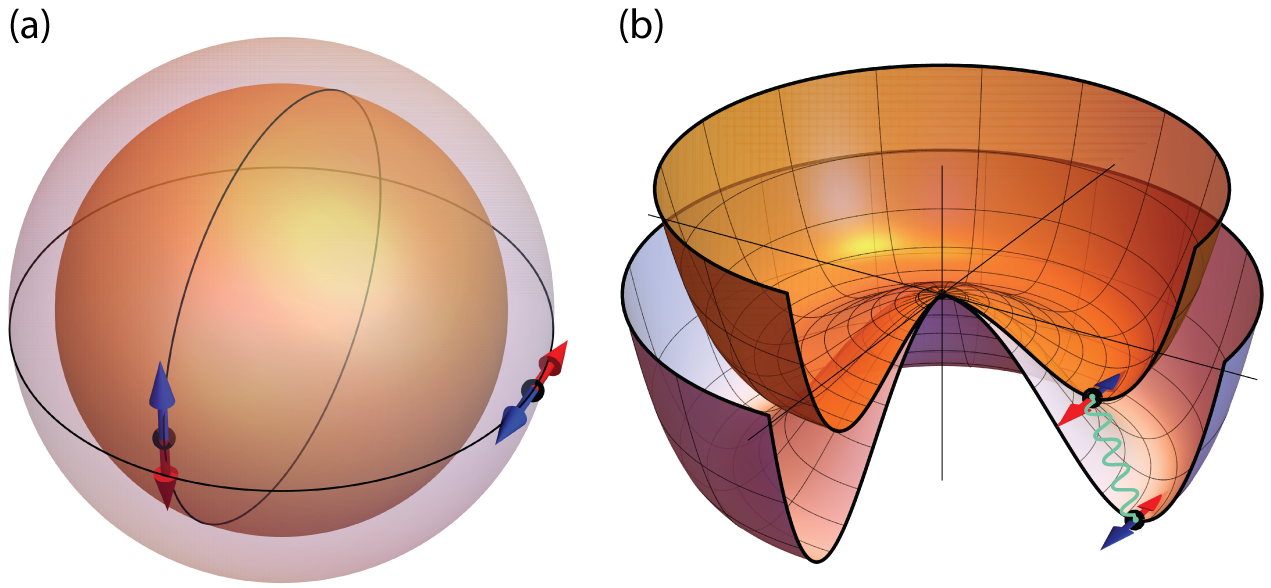}
\caption{Illustration of the Leggett phase mode for $Fdd2$ with (a) the 
free energy surfaces drawn as nested spheres for $\Gamma_4^-$ and $\Gamma_5^-$, whose radii are the amplitudes $Q_4$ and $Q_5$, and 
(b) the corresponding Mexican-hat potentials when restricting to $Cc$.  
In (a) the $Cc$ subspace is indicated by great circles on these spheres (which are in reality warped given that $Q_4$ and $Q_5$ depend on the spherical angles). The $Fdd2$ minimum in Cartesian coordinates is $(Q_4,0,0)$ on the $\Gamma_4^-$ sphere and $(0,Q_5,0)$ on the $\Gamma_5^-$ sphere.  
In (b) the vertical axis is energy, with the wavy line indicating the Landau couplings between the two energy surfaces.
In both panels, the Goldstone mode corresponds to oscillations about the $Fdd2$ minima on each circle (these minima are indicated by black dots) to one $Cc$ domain on one swing (a red arrow on one surface, a blue arrow on the other) and to another $Cc$ domain on the other swing (a blue arrow on one surface, a red arrow on the other).  
The higher energy Leggett mode (anti-phason) instead involves  oscillations corresponding to antiphase $Cc$ domains, with one swing
involving both red arrows and the other swing involving both blue arrows.  A similar picture exists for the antiphase Higgs mode,
with the arrows pointing along the radial directions instead.}
\label{fig1}
\end{figure*}

What about the anti-phason, the out-of-phase mode in the phase channel that would be the analog of the superconducting Leggett mode?  To obtain this requires coupling of the two Goldstone modes.  One can show that such coupling requires terms in \autoref{eq:bq}  that are linear in the two Goldstone variables. In the above example, this would be terms of the form $q_2 r_2$ times Higgs variables ($q_1, r_1, q_1^2, r_1^2, q_1 r_1$, etc.).  Although these terms do not exist in
the above minimal model, they do exist in general.  Instead of extending our minimal model to the more general case, we instead illustrate anti-phasons using the specific example of Cd$_2$Nb$_2$O$_7$.

\section{Landau Theory of C\lowercase{d}$_2$N\lowercase{b}$_2$O$_7$}

The pyrochlore Cd$_2$Nb$_2$O$_7$ is one of the few known stoichiometric materials that exhibits relaxor ferroelectric behavior.  Because it is stoichiometric, its phonons, as observed by IR reflectivity and Raman scattering, are well defined.  The pyrochlore structure consists of a network of corner sharing NbO$_6$ octahedra interpenetrated by CdO tetahedra (or CdO chains depending on how one views it), as shown in \autoref{figx}. Several structural transitions have been observed as a function of temperature, where the symmetry lowers from the high temperature cubic phase ($Fd\bar{3}m$).  Of particular interest are the ferroelastic transition at 204 K (which lowers the point group symmetry from $m\bar{3}m$ to $mmm$, maintaining inversion) and  the ferroelectric one at 196 K (which further lowers the point group symmetry to $mm2$) \cite{ye}.  At much lower temperatures, two other transitions have been reported that are consistent with a monoclinic space group, probably $Cc$.  Based on group-subgroup relations and structural refinements, the accepted phase transition series with decreasing temperature is $Fd\bar{3}m \rightarrow Imma \rightarrow Ima2 \rightarrow Cc$.  The complexity of the phase diagram and the fact that the polarization direction is easily reoriented by an external field \cite{ye} indicate that the free energy landscape is soft, suggesting that this material is an excellent hunting ground for new collective modes.

Density functional theory (DFT) calculations ($T=0$\,K) show that $Fd\bar{3}m$ is unstable to both $\Gamma_4^-$ and $\Gamma_5^-$ distortions (the former being polar in nature).
$\Gamma_4^-$ and $\Gamma_5^-$ are both primary
order parameters for the transition from $Fd\bar{3}m$ to $Ima2$
\footnote{The speculated $Imma$ ferroelastic phase exists over a very limited range in temperature and its nature is ambiguous; if Cd$_2$Nb$_2$O$_7$ does experimentally adopt $Imma$ symmetry, then it would be driven by a  $\Gamma_5^+$ distortion which is found not to occur in DFT studies \cite{fischer,laurita}.}.
In the ferroelectric phase, the minimum Landau subspace is  $6\times6$, given that both $\Gamma_4^-$ and $\Gamma_5^-$ are three-dimensional group representations, meaning Cd$_2$Nb$_2$O$_7$ hosts a richer space of possibilities than the minimal Landau model discussed earlier \cite{Talanov2021}.
In addition, as we demonstrate below, more coupling terms exist besides the biquadratic one, a general result not specific to Cd$_2$Nb$_2$O$_7$.  Before turning to $Ima2$, we first discuss the related $Fdd2$ space group which is simpler to present in a Cartesian basis.

\begin{table}
\caption{\label{table1}Space groups generated on condensing the ($\Gamma_4^-$, $\Gamma_5^-$) order parameters along various crystallographic directions in Cd$_2$Nb$_2$O$_7$ (generated from Ref.~\onlinecite{byu}).}
\begin{ruledtabular}
\begin{tabular}{lll}
$\Gamma_4^-$  & $\Gamma_5^-$ & Space Group \\
\hline
(a,0,0) & (0,b,0) & $Fdd2$ \\
(a,a,0) & (0,b,-b) & $Ima2$ \\
(a,a,a) & (b,b,b) & $R3$ \\
(a,b,0) & (0,c,d) & $Cc$ \\
(a,a,b) & (0,c,-c) & $Cm$ \\
(a,a,0) & (-c,b,-b) & $C2$ \\
\end{tabular}
\end{ruledtabular}
\end{table}

\subsection{$Fdd2$}
The six-dimensional subspace formed by $\Gamma_4^-$ and $\Gamma_5^-$ defines two free energy surfaces (\autoref{fig1}a).  These surfaces are three-dimensional generalizations of two Mexican-hat potentials (\autoref{fig1}b), and can be considered as nested warped spheres, one for $\Gamma_4^-$, the other for $\Gamma_5^-$. Each representation condensing along the cubic axis of its respective sphere gives rise to $Fdd2$. Interestingly, $Fdd2$ has been identified as the local structure of Cd$_2$Nb$_2$O$_7$ from diffuse scattering studies \cite{malcherek2} and is also the global space group upon sulfur doping \cite{laurita}.  But what is evident from both the DFT and structural studies is that there are a number of space groups which are close in energy and provide comparable descriptions of the data.  That is, the free energy landscape of these two coupled spheres is relatively flat (i.e., the warping of the spheres is small).  DFT calculations indicate that structures where $\Gamma_4^-$ is dominant have a slightly lower free energy than ones where $\Gamma_5^-$ is dominant \cite{laurita}, whereas order parameter-like behavior of the distortion amplitudes as a function of temperature has only been claimed for $\Gamma_5^-$ based on global structural refinements \cite{malcherek}.  Regardless, the net result is that one has two coupled flat energy surfaces, one for $\Gamma_4^-$, the other for $\Gamma_5^-$, which only differ from each other by a few meV per formula unit \cite{laurita}.

In terms of order parameter directions (\autoref{table1}), the $Fdd2$ phase corresponds to $(a,0,0)|(0,c,0)$ with $(a,0,0)$ being a point on the $\Gamma_4^-$ sphere and $(0,c,0)$ on the $\Gamma_5^-$ sphere, whereas $Ima2$ corresponds to $(a,a,0)|(0,c,-c)$ instead.  Here, $a$ refers to the $\Gamma_4^-$ order parameter and $c$ to the
$\Gamma_5^-$ one.  The lower symmetry space group encompassing these two is $Cc$, corresponding to $(a,b,0)|(0,c,d)$, which can be seen to correspond to particular great circles on each sphere (analogous to the Mexican-hat brims of the minimal model).  For our purposes here,
we will restrict our analysis to these two circles, recognizing that there are also fluctuations orthogonal to them on the spheres corresponding to rhombohedral ($R3$) and other monoclinic ($Cm$ and $C2$) space groups, as well as to other domains of $Fdd2$ and $Ima2$ not represented by these two particular circles   (\autoref{table2}).  That is, we consider a reduced $4\times4$ force-constant matrix.
We note that for these flat free energy surfaces, it is often useful to describe each order parameter with spherical/polar coordinates, but for simplicity in the following derivations we will use a Cartesian basis instead.

We obtain the Landau free energy using the INVARIANTS routine \cite{inv} (accessible electronically at Ref.~\onlinecite{byu}), noting that it is important to specify the appropriate $Cc$ domains for the subspace considered, specifically the order parameter directions $(a,b,0)|(0,c,d)$. 
Considering terms to quartic order, we obtain
\begin{eqnarray}
F &  = & \frac{\alpha_4}{2}(a^2+b^2)+\frac{\alpha_5}{2}(c^2+d^2)+\frac{\beta_4}{4}(a^2+b^2)^2 \nonumber \\
& &  +\frac{\gamma_4}{4}(a^4+b^4)+\frac{\beta_5}{4}(c^2+d^2)^2+\frac{\gamma_5}{4}(c^4+d^4) \nonumber \\
& & +\frac{\delta}{2}(a^2+b^2)(c^2+d^2)+\epsilon_1(ab)(ad-bc) \nonumber \\
& & +\frac{\epsilon_2}{2}(a^2c^2+b^2d^2) +\epsilon_3(abcd) +\epsilon_4(cd)(ad-bc)\,. \nonumber \\
\label{eq:landauCNO}
\end{eqnarray}
We note that in polar coordinates, $a=Q_4\cos(\phi_4), b=Q_4\sin(\phi_4), c=Q_5\cos(\phi_5), d=Q_5\sin(\phi_5)$, where $Q_i$ are the amplitudes and $\phi_i$ are the phases.  

We first consider expanding about an assumed $Fdd2$ free energy minimum, and then consider $Ima2$ next.  For $Fdd2$, fluctuations of $a$ and $c$ correspond to Higgs modes ($A_1$ symmetry), fluctuations of $b$ and $d$ to Goldstone modes ($B_2$ symmetry) (\autoref{table2}).  Note these Goldstone modes are not at zero energy because of the anisotropy terms $\gamma_i$, so they are formally pseudo-Goldstone modes.  In addition, the $\delta$ (biquadratic) and $\epsilon_i$ terms will provide coupling between the two Higgs modes, and the latter also between the two Goldstone modes.  There is no coupling between the Higgs and the Goldstone modes, so the 4 $\times$ 4 matrix reduces to two  $2\times2$  blocks: one for the two Higgs modes, one for the two Goldstone modes.
This remains true when considering sixth-order terms in the free energy, and also for the full $\Gamma_4^- \oplus \Gamma_5^-$ space.  That is, this larger  $6\times6$ matrix reduces to three $2\times2$ blocks: one in the Higgs sector and the other two in the Goldstone sector.
That is, there are no terms coupling the two Goldstone blocks.

Differentiating \autoref{eq:landauCNO} with respect to $a,b,c,d$ and setting each expression to zero determines these parameters via a set of coupled equations \cite{holo}.  One can see from these expressions that $b=d=0$ ($Fdd2$) is an allowed solution to these coupled equations.
Whether it is or is not depends on the actual values of the Landau coefficients, but for purposes here we assume that this is the case 
\footnote{The $a,b,c,d$ coefficients can be determined from DFT \cite{artyukhin}, or else extracted from an AMPLIMODES \cite{amp,bilbao1,bilbao2} or ISODISTORT \cite{iso,byu} analysis of the experimental structural refinements.}.

The coefficients of the force-constant matrix for each of the modes are determined by $\frac{\partial^2F}{\partial q_i \partial q_j}$ where $q_i$ are $a,b,c,d$, evaluated at $a=\left<a\right>$, $c=\left<c\right>$, $b=d=0$  The diagonal elements of the force-constant matrix will give the uncoupled Higgs mode energies ($aa$ and $cc$ elements) and Goldstone mode energies ($bb$ and $dd$ elements).  Of interest here are the off-diagonal terms. We find that in this case, there are two non-zero ones. The first is the $ac$ one that couples the two Higgs modes
\begin{equation}
\frac{\partial^2F}{\partial a \partial c} = 2(\delta+\epsilon_2)(ac) \, .
\end{equation}
The other is the $bd$ one that couples the two Goldstone modes
\begin{equation}
\frac{\partial^2F}{\partial b \partial d} = \epsilon_1(a^2) + \epsilon_3 (ac) -\epsilon_4(c^2) \, .
\end{equation}
For each 2 $\times$ 2 block, we denote the diagonal elements corresponding to the uncoupled mode energies as $\omega_4^2$ ($aa$ or $bb$) and  $\omega_5^2$ ($cc$ or $dd$).  Denoting the off-diagonal element in each block as $X$ ($ab$ or $cd$), one obtains as before coupled mode energies of the form
\begin{equation}
\omega_{\pm}^2 = \frac{\omega_4^2+\omega_5^2}{2} \pm \sqrt{\frac{(\omega_4^2-\omega_5^2)^2}{4}+X^2}\,.
\end{equation}
The two eigenvectors of this matrix either have the two components in-phase, or out-of-phase, as in \autoref{eq:eigs}. 
The in-phase modes then correspond to Higgs and Goldstone modes, the out-of-phase modes to their Leggett analogs.

To understand the two phase modes, consider a given $Fdd2$ domain, $(a,0,0)|(0,c,0)$. The Goldstone mode would correspond to oscillations along the two circles on the free energy spheres towards one $Cc$ domain $(a,b,0)|(0,c,d)$
(one swing) and towards another $Cc$ domain $(a,-b,0)|(0,c,-d)$ (the other swing) as illustrated in \autoref{fig1}.  The Leggett analog corresponds instead to oscillations towards $(a,b,0)|(0,c,-d)$
and $(a,-b,0)|(0,c,d)$.  These can be thought of as antiphase domains of $Cc$ that occur at a higher energy since they are
penalized by the coupling terms in the Landau free energy.  As such,
the Leggett phase mode occurs at a higher energy than the Goldstone mode.  We therefore denote the Leggett phase mode as an anti-phason.  A similar picture applies in the Higgs sector (oscillations along the radial directions in \autoref{fig1}).  That is, the amplitudes $Q_4$ and $Q_5$ oscillate either in-phase or out-of phase, so we can denote the latter as a Leggett amplitude mode, that is an antiphase Higgs mode.  Considering the full six-dimensional $\Gamma_4^- \oplus \Gamma_5^-$ space, there are actually two `anti-phasons' 
and one `antiphase Higgs' mode, as alluded to above.

\subsection{$Ima2$}
Although the $Fdd2$ space group is realized upon sulfur doping \cite{laurita}, 
stoichiometric Cd$_2$Re$_2$O$_7$ is thought to be $Ima2$ below the ferroelectric transition.  The $Ima2$ case is similar to the $Fdd2$.  $Ima2$ corresponds to $a=b$ and $d=-c$.  These coordinates represent points on the two circles in \autoref{fig1} that are rotated by 45$^\circ$ relative to $Fdd2$.  Therefore, to construct Goldstone and Higgs variables in a Cartesian basis, it is convenient to express $F$ using a 45$^\circ$ rotated coordinate frame instead, that is $\tilde{a}=(a+b)/\sqrt{2},\, \tilde{b}=(a-b)/\sqrt{2},\, \tilde{c}=(c-d)/\sqrt{2},\, \tilde{d}=(c+d)/\sqrt{2}$, so that $Ima2$ corresponds to $\tilde{b}=\tilde{d}=0$.  The resulting Landau free energy is:
\begin{eqnarray}
F & = & \frac{\alpha_4}{2}(\tilde{a}^2+\tilde{b}^2)+\frac{\alpha_5}{2}(\tilde{c}^2+\tilde{d}^2)+\frac{\beta_4}{4}(\tilde{a}^2+\tilde{b}^2)^2 \nonumber \\
& & +\frac{\gamma_4}{8}(\tilde{a}^4+\tilde{b}^4+6\tilde{a}^2\tilde{b}^2)+\frac{\beta_5}{4}(\tilde{c}^2+\tilde{d}^2)^2 \nonumber \\
& & +\frac{\gamma_5}{8}(\tilde{c}^4+\tilde{d}^4+6\tilde{c}^2\tilde{d}^2) +\frac{\delta}{2}(\tilde{a}^2+\tilde{b}^2)(\tilde{c}^2+\tilde{d}^2) \nonumber \\
& & +\frac{\epsilon_1}{2}(\tilde{a}^2-\tilde{b}^2)(\tilde{b}\tilde{d}-\tilde{a}\tilde{c}) \nonumber \\
& & +\frac{\epsilon_2}{4}[(\tilde{a}^2+\tilde{b}^2)(\tilde{c}^2+\tilde{d}^2) +4\tilde{a}\tilde{b}\tilde{c}\tilde{d}] \nonumber \\
& & +\frac{\epsilon_3}{4}(\tilde{a}^2-\tilde{b}^2)(\tilde{d}^2-\tilde{c}^2) +\frac{\epsilon_4}{2}(\tilde{d}^2-\tilde{c}^2)(\tilde{b}\tilde{d}-\tilde{a}\tilde{c})\,. \nonumber \\
\end{eqnarray}
One can show that, as before, the only off-diagonal terms of the force-constant matrix that survive are the $\tilde{a}\tilde{c}$ and $\tilde{b}\tilde{d}$ terms.  Thus, the force-constant matrix again reduces to two $2\times2$  blocks and results in a Higgs mode, a Goldstone mode, and the two Leggett modes: an antiphase Higgs mode and an anti-phason.
This is straightforward to understand from a study of \autoref{fig1} and \autoref{table1} as we summarize in \autoref{table2}.

\begin{table}
\caption{Fluctuations about $Fdd2$ and $Ima2$ derived from Table I.  H denotes fluctuations indicated by the blue/red arrows along the specific great circles shown on the spheres in \autoref{fig1}.  V denotes fluctuations along great circles orthogonal to these circles. Note that the circles are turned 90 degrees between the two surfaces because of the different transformation properties of $\Gamma_4^-$ and $\Gamma_5^-$. R denotes fluctuations along the radial directions of the spheres.  For $Fdd2$, the two $Cc$ rows correspond to different $Cc$ domains.  Also, the reduction to $C2$ is not shown because it involves an $Fdd2 \rightarrow P1 \rightarrow C2$ path on the spheres in \autoref{fig1}.}
\begin{ruledtabular}
\begin{tabular}{lllll}
Space group & Symmetry & $\Gamma_4^-$ & $\Gamma_5^-$ & Modes  \\
\hline
$Fdd2 \rightarrow Fdd2$ & $\Gamma_1$ (A$_1$) & R & R & Higgs, antiphase Higgs \\
$Fdd2 \rightarrow Cc$ & $\Gamma_4$ (B$_2$) & H & H & Goldstone, anti-phason \\
$Fdd2 \rightarrow Cc$ & $\Gamma_3$ (A$_2$) & V & V & Goldstone, anti-phason \\
\hline
$Ima2 \rightarrow Ima2$ & $\Gamma_1$ (A$_1$) & R & R & Higgs, antiphase Higgs \\
$Ima2 \rightarrow Cc$ & $\Gamma_4$ (B$_2$) & H & H & Goldstone, anti-phason \\
$Ima2 \rightarrow Cm$ & $\Gamma_3$ (A$_2$) & V & & Goldstone \\
$Ima2 \rightarrow C2$ & $\Gamma_2$ (B$_1$) & & V & Goldstone \\
\end{tabular}
\end{ruledtabular}
\label{table2}
\end{table}

We now connect these results to those on superconductors.
Upon manipulation of the corresponding dynamical matrix in the case of superconductivity \cite{sharapov}, one can show that the off-diagonal element for a charge-neutral two-band superconductor is equal to the square root of the product of the two diagonal elements.  This results in one mode frequency that is zero (the Goldstone mode) and one whose squared frequency is the sum of the two diagonal elements (the Leggett mode).  Besides this distinction, there is no qualitative difference between the superconductivity and structural mode cases. In particular, for both cases, the two components of the eigenvector do not have equal amplitudes.  In the superconductivity case, this is due to the difference in the density of states of the two bands.  In the structural case, it is because the uncoupled mode frequencies for $\Gamma_4^-$ and $\Gamma_5^-$ differ.

The Leggett analog of the Higgs mode (the antiphase Higgs) has not been explicitly described before. A related mode, however, was predicted recently in time-reversal breaking superconductors \cite{prineha}.  It should be observable as well in multi-band superconductors, and would correspond to out-of-phase oscillations of the gap amplitudes of the two bands. We suggest that the existence of this mode be searched for by appropriate experiments (Raman, pump-probe) on two-band superconductors like MgB$_2$.  Below, we address its observation in the structural case.

\subsection{Secondary Order Parameters}
The secondary order parameters present in  Cd$_2$Nb$_2$O$_7$ are even-parity ones with symmetry $\Gamma_3^+$ and $\Gamma_5^+$.  For $Cc$, this results in a seven-dimensional space:  $\Gamma_4^- (a,b) \oplus \Gamma_5^- (c,d)\oplus \Gamma_3^+ (e,f)\oplus \Gamma_5^+ (g)$.  Expanding around $Fdd2$ ($b=d=g=0$) and $Ima2$ ($b=a, d=-c, f=0$), no couplings exist between the primary Higgs and Goldstone modes (that is, the $ab, ad, bc, cd$ terms in the force-constant matrix vanish at the $Fdd2$ and $Ima2$ minima).  This means that any coupling between these Higgs and Goldstone modes are beyond the harmonic approximation, however higher-order coupling can occur in non-equilibrium situations \cite{meier, juraschek}. At first glance, these secondary modes shift the various primary mode frequencies that would be determined from the smaller $4\times4$ block.  That is, one can in principle reduce this $7\times7$ matrix to an effective  $4\times4$ one by integrating out $e,f,g$ in the Landau free energy equations, as these secondary order parameters are slaved to the primary ones.

But closer inspection of the form of the primary-secondary mode couplings reveals new opportunities to study the Leggett modes.  To see this, note that in the $Cc$ subspace discussed above, one now has the following coupling terms at cubic order since the secondary order parameters have even parity: 
\begin{eqnarray}
F_3 & = & \eta_1 [\frac{1}{\sqrt{3}}(a^2+b^2)e + (a^2-b^2)f] \nonumber \\
& & +\eta_2 [\frac{1}{\sqrt{3}}(c^2+d^2)e + (c^2-d^2)f] \nonumber \\
& & +\eta_3 [(ac-bd)e-\frac{1}{\sqrt{3}}(ac+bd)f] \nonumber \\
& & +\eta_4 (abg) +\eta_5 (cdg) +\eta_6 [(ad-bc)g]\,.
\end{eqnarray}
For both $Fdd2$ and $Ima2$, the $\eta_3$ term leads to both primary Higgs and Goldstone mode couplings.
For $Ima2$, the $\eta_6$ term also leads to both primary Higgs and Goldstone couplings.
We note that $\Gamma_3^+$ and $\Gamma_5^+$ are Raman active modes in the cubic phase.  Therefore, below
the ferroelectric transition, they can be used to drive the primary Leggett modes via these two cubic terms, analogous
to nonlinear phononics experiments on perovskites \cite{forst} that have been studied theoretically using similar Landau-like equations of motion
\cite{subedi,juraschek2,gu}.  Note that pump-probe experiments have been instrumental in
the study of Leggett modes in superconductors \cite{giorgianni}.

In the above analysis, we did not include strain (which typically renormalizes the Landau coefficients), and gradient terms which need to be included to address domain walls.  One would expect that the collective modes could be significantly modified by domain walls if they are spatially broad enough \cite{huang} and present at a high enough density.  These effects would be interesting to study in future work.

\begin{figure}
\centering
\includegraphics[width=0.95\columnwidth]{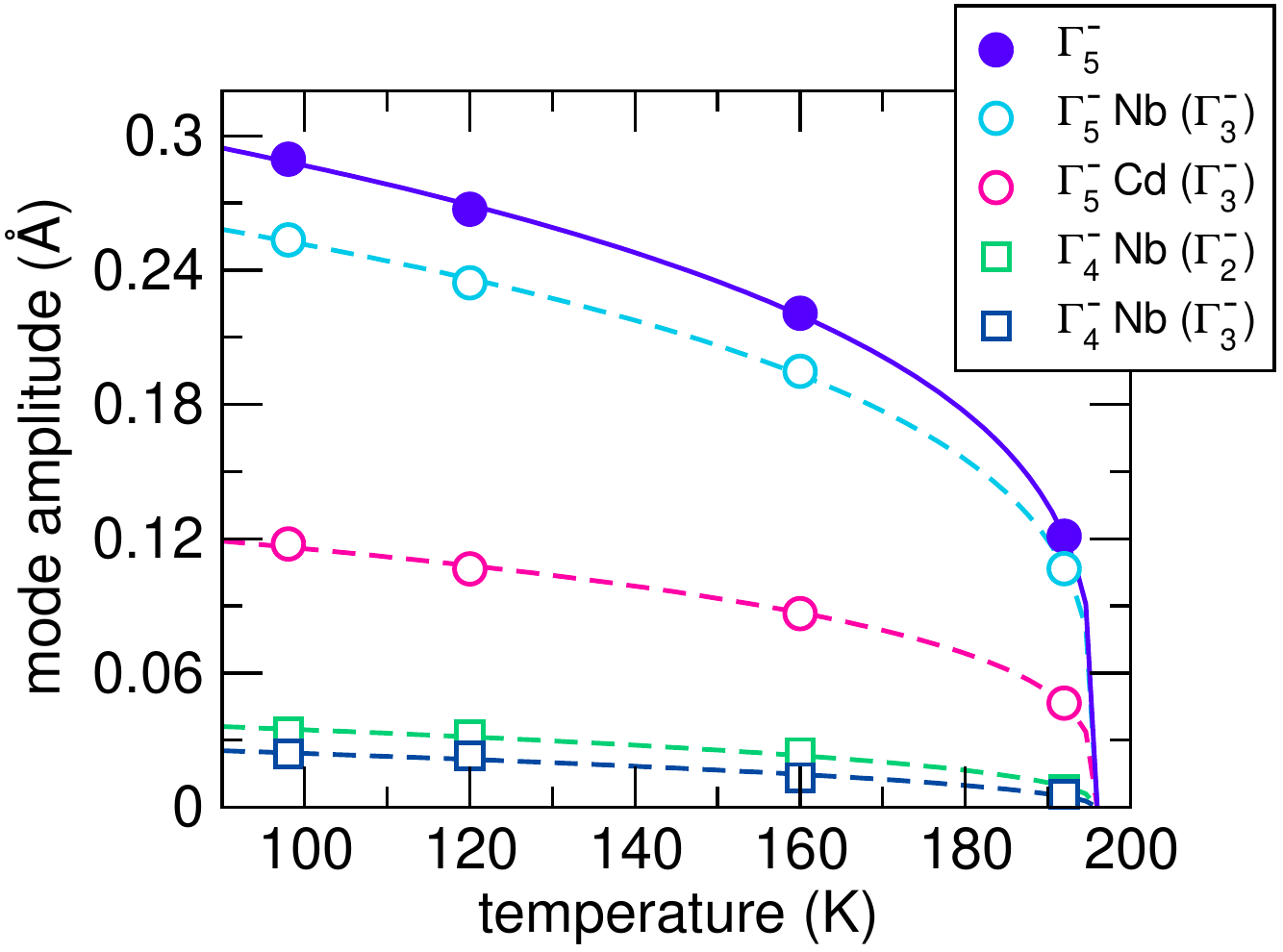}
\caption{Decomposition of the $Ima2$ structural refinement of Cd$_2$Nb$_2$O$_7$ from Ref.\ \onlinecite{malcherek} using ISODISTORT \cite{iso,byu}.  Shown is the overall $\Gamma_5^-$ ($T_{2u}$) amplitude previously plotted in \cite{malcherek} along with various $\Gamma_4^-$ ($T_{1u}$) and $\Gamma_5^-$ decompositions involving either Nb or Cd ion displacements from their cubic positions, noting that for each ion, $\Gamma_4^-$ has two submodes $\Gamma_3^-$ ($E_u$) and $\Gamma_2^-$ (A$_{2u}$) whereas $\Gamma_5^-$ has only one ($\Gamma_3^-$).  Curves are proportional to $|T_c-T|^\beta$,
where $T_c=196$\,K is the ferroelectric transition temperature and $\beta=0.27$ is the order parameter exponent.}
\label{fig2}
\end{figure}

\subsection{Submodes and Phonons for Cd$_2$Nb$_2$O$_7$}
In the real crystal, there are multiple force-constant matrix eigenmodes for each symmetry, typically referred to in the literature as `submodes'. The full matrix has a size of $66\times66$.  Restricting the symmetry to $Cc$, the matrix reduces to $33\times33$.  Further restriction to $\Gamma_4^-$ and $\Gamma_5^-$, it reduces to 24 $\times$ 24.
In the Landau treatment, each $q_i$ is a particular sum of these submodes with the appropriate group symmetry that is gotten
by reducing this larger matrix to the smaller $4\times4$ matrix.  Still, one can ask the question whether this simplification is supported by the data or not.
To address this, in \autoref{fig2}, we plot the temperature dependence of several relevant submode amplitudes based on the $Ima2$ crystal structure refinements  of Malcherek {\it et al.}~\cite{malcherek}.  We find  that the Cd and Nb displacements for both $\Gamma_4^-$ and $\Gamma_5^-$ symmetries scale with the overall $\Gamma_5^-$ order parameter amplitude that was previously shown in Ref.\ \cite{malcherek}. 
This indicates that this simpler Landau description is valid.

We now turn to the phonons.  Since the ions involved (Cd, Nb, O) have different masses, there is no one-to-one correspondence between the force-constant modes and the phonons (unless the mode involved only one ion type).  As a consequence, a given force-constant mode could involve more than one phonon. However, it has been shown that there is a strong correspondence between single phonons and the Higgs and Goldstone modes in YMnO$_3$ \cite{meier}.  Therefore it is probable that such modes for Cd$_2$Nb$_2$O$_7$ can also be associated with specific phonons.

To gain further insight, we turn to experimental Raman and infrared (IR) data.
In the $Ima2$ phase, all phonons are in principle Raman active.
Raman data on Cd$_2$Nb$_2$O$_7$ find several low lying $A_1$ and $B_2$ modes below the ferroelectric transition, three of which soften as the structural phase transition is approached from below \cite{taniguchi}.  As the amplitude modes have $A_1$ symmetry and the phase modes have $B_2$ symmetry, there should be a correlation between our collective modes and the data.  That is, it is possible that the two lowest lying $A_1$ Raman modes corresponds to the Higgs and antiphase Higgs modes.  Presumably the lowest lying $B_2$ mode is the Goldstone mode, with the anti-phason corresponding to a higher energy $B_2$ mode that has yet to be studied.  

The IR data for Cd$_2$Nb$_2$O$_7$  are complicated given the presence of seven optic modes of $\Gamma_4^-$ symmetry in the cubic phase, which further split in the ferroelectric phase.  Two of the lower lying IR modes have been interpreted as being coupled (both above and below the transition) due to their temperature dependencies  \cite{buix}.  One of these, referred to as a `central' mode and speculated to be due to hopping of Cd ions among equivalent locations displaced from their cubic positions, is not thought to be of $\Gamma_4^-$ symmetry given that seven other IR modes of this symmetry are already seen. Its coupling with a higher energy second mode, thought to be a Nb displacement mode of $\Gamma_4^-$ symmetry, drives this central mode to soften at the ferroelectric transition \cite{buix}. Whether this central mode is an $A_1$ symmetry mode (as typical for an order-disorder transition), or instead a $\Gamma_5^-$ mode that becomes IR active due to coupling to the second $\Gamma_4^-$ mode, is worth investigating.  If the latter, this would be consistent with the theory we offer above of two coupled modes of $\Gamma_4^-$ and $\Gamma_5^-$ symmetry.  Moreover, as discussed above, pump-probe studies of Cd$_2$Nb$_2$O$_7$ would
be instrumental in probing for possible collective modes: Higgs, Goldstone, and their Leggett analogs.

\section{Searching for and Observing Leggett Modes}

Although we considered the specific example of Cd$_2$Nb$_2$O$_7$, the above analysis should be applicable whenever more than one primary order parameter is involved in a phase transition.
As such, Leggett modes should exist in the structural context, and could be identified from a DFT analysis of the force-constant and dynamical (phonon) matrices in comparison to experimental data.  We plan to report on this in a future paper that will provide a detailed DFT study of Cd$_2$Nb$_2$O$_7$ \cite{young}.  In general, analysis of Raman and IR data, including the symmetry of the modes and their temperature dependence, should be helpful in elucidating the presence of Leggett modes.  Specifics, including
identifying the out-of-phase behavior characteristic of the Leggett modes, could be resolved by pump-probe studies of the phase of the oscillations as a function of time.  Moreover, the
equations of motion associated with nonlinear phononics involve the same cubic and quartic coupling
terms invoked in this paper that provide for the existence of the Leggett modes to begin with.  Driving specific modes would then be instrumental in identifying the various collective modes via their couplings to the driven mode.

As for materials, obviously those phase transitions involving more than one primary group representation would be obvious
targets for study, including those exhibiting improper and hybrid-improper ferroelectric transitions \cite{bousquet,xu,bellaiche,benedek,Mulder2013,perez-mato}.  Although the latter typically involve order parameters with non-zero wavevector, the Landau coupling terms in the free energy are similar and so the considerations presented here should be valid there as well.  The existence of low lying modes would be aided by having flat energy surfaces with low transition barriers as considered here.  In that context, Cd$_2$Nb$_2$O$_7$ consists of corner sharing NbO$_6$
octahedra in an open framework interpenetrated by CdO tetrahedra that are weakly coupled to the octahedra, implying
floppy low energy modes. Going from a cubic phase to a lower symmetry orthorhombic or monoclinic phase
is also useful in order to optimize the number of coupling terms in the Landau free energy.  This is
particularly pervasive in pyrochlores and spinels.  Cage-like structures as in skutterudites with their associated floppy modes would also be a good place to search.  Much of the work concerning Goldstone and Higgs modes has been done in the context of perovskites, which also involve corner-sharing octahedra, and a number of them exhibit improper ferroelectric transitions as referenced above.
Similar considerations to ours would also apply to ferroelastic transitions. However, a large coupling to strain usually leads to large warping terms (see e.g.\ in the case of ferroelastic WO$_3$ \cite{mascello}). A locally flat energy landscape can sometimes be recovered at Ising-type domain walls (see e.g.\ for LiNbO$_3$ \cite{Mukherjee}). In the same way, locally-confined Leggett modes could be observed.

\section{Conclusion}
We demonstrated via a Landau analysis that structural Leggett modes should exist in the context of displacive transitions when more than one multi-dimensional group representation is involved.
These collective modes exist not only in the phase channel as in the original work of Leggett, but also in the amplitude channel, representing a new collective mode, the antiphase Higgs, that should be observable in multi-band superconductors as well.
We studied in detail the specific case of the relaxor ferroelectric pyrochlore Cd$_2$Nb$_2$O$_7$, which we believe is a promising material to search for such modes.  We believe there should be a large group of materials where such modes could exist, and advocate in particular that
pump-probe studies would be the most illuminating way to identify and characterize these modes.
The study of such modes should give new insights into their associated crystallographic phase transitions.

\begin{acknowledgments}
Work at Argonne National Laboratory was supported by the Materials Sciences and Engineering
Division, Basic Energy Sciences, Office of Science, U.S.~Dept.~of Energy. Q.N.M.\ was supported by the Swiss National Science Foundation under project number P2EZP2\_191872.
Work at ETH Zurich was funded by the European Research Council (ERC) under the European Union’s Horizon 2020 research and innovation program project HERO grant (No.~810451).
Work at Northwestern University was supported by the National Science Foundation (NSF) Grant No.\ DMR-2011208. G.L.\ gratefully acknowledges support for this work from Bates College, and from NSF through DMR-1904980.
\end{acknowledgments}

\bibliography{references}

\end{document}